\documentclass[twocolumn,prb,floatfix,showpacs]{revtex4-1}

\usepackage{graphicx}
\usepackage{bm}

\begin{document}
\title{Isolated Dirac cone on the surface of ternary tetradymite-like
topological insulators}

\author{H. Lin$^1$, Tanmoy Das$^{1,2}$, L. A. Wray$^3$, S.-Y. Xu$^3$, M. Z.
Hasan$^{3}$, and A. Bansil$^1$ }

\affiliation{ $^1$Department of Physics, Northeastern University,
Boston, MA 02115, USA\\
 $^2$Theoretical Division, Los Alamos National Laboratory, Los Alamos,
NM 87545, USA\\
 $^3$ Joseph Henry Laboratories of Physics, Princeton University, Princeton,
New Jersey 08544, USA\\}


\begin{abstract}
We have extended our search for topological insulators to the
ternary tetradymite-like compounds $M_2X_2Y$ ($M$=Bi and Sb; $X$ and
$Y$ = S, Se, and Te), which
are variations of the well-known binary compounds Bi$_2$Se$_3$ and Bi$_2$Te$_3$.
Our first-principles computations suggest
that five existing compounds are strong topological insulators with
a single Dirac cone on the surface.
In particular, stoichiometric Bi$_2$Se$_2$S, Sb$_2$Te$_2$Se, and
Sb$_2$Te$_2$S are predicted to have an
isolated Dirac cone on their naturally cleaved surface.
This finding paves the way for realizing the topological transport regime.
\end{abstract}

\maketitle
Topological insulator is a very distinct state of quantum matter with
a bulk-insulating gap of spin-orbit coupling (SOC) origin.
\cite{review, reviewMZCL,
Kane1st,FuKane,DavidNat1,KonigSci,DavidTunable,TIbasic,BernevigSciHgTe,Roy,MooreandBal}
The Dirac fermions on the surface of a topological insulator possess
topological quantum Berry phase and a definite spin chirality, which
protects them from back-scattering and localization, resulting in
counterpropagation of opposite spin states without dissipation.
These are the key electronic ingredients for applications to
topological quantum computing\cite{Majorana,LeekQC} and low-power
spintronics\cite{WolfSpintronics} devices for which,
however, the dissipationless surface states must be in the topological
transport regime, i.e. to have an isolated Dirac cone fully separated
from the bulk bands and the Fermi level to lie at the Dirac
point.\cite{KaneDevice}
The existing topological insulating materials fail in this regard
due to their material-specific complications such as
interference from numerous surface states\cite{ZhangPred}, shielding
of the Dirac cone either inside the bulk-valence
band\cite{MatthewNatPhys,BiTeSbTe} or the conduction
bands\cite{TlBiTe2}, or the involvement of
crystal distortion in fabricating the material
\cite{heuslerhasan,HsinLi}.

Recently, it was shown experimentally
that an isolated Dirac cone can be achieved by tuning the Fermi
level with appropriate hole doping, but the non-stoichiometric
crystal structure and the requirement of surface deposition generally
makes this approach
unsuitable for most practical applications.\cite{DavidTunable,BiTeSbTe}
On the other hand, solid solutions in the tetradymite-like compounds
provide a platform for engineering topological surface states.
For example, theoretical work of Zhang et al.\cite{ZhangNJP}
indicates the presence of a topological phase
transition in Sb$_2$(Te$_{1-x}$Se$_x$)$_3$ for $0 \leq x \leq 1$.
Interestingly, since Se atoms preferentially occupy the central site
in the crystal,
at $x$=1/3 the stoichiometric compound Sb$_2$Te$_2$Se can admit an
ordered phase.
Our recent work on topological insulator Bi$_2$Te$_2$Se reveals
that the linewidth of the topological surface states in
angle-resolved photoemission spectroscopy (ARPES) is narrow,
indicating that disorder effects are suppressed in these
compounds.\cite{SuYangGBT}
Ren et al.\cite{BTS} have shown that Bi$_2$Te$_2$Se is the best
material to date for studying the topological surface states due to
its highest bulk resistivity among all known topological insulators.

Here, we report first-principles computations on five existing
tetradymite-like compounds Bi$_2$Te$_2$Se, Bi$_2$Te$_2$S,
Bi$_2$Se$_2$S, Sb$_2$Te$_2$Se and Sb$_2$Te$_2$S to show that these
materials host salient topological insulating features in their
stoichiometric crystal structures. Our important findings
include that the last three of the five aforementioned compounds harbor an
isolated Dirac cone with Dirac points which lie fully within the gap, allowing
these states to
reach the long-sought territory of the topological
spin-transport regime\cite{TIbasic,ZhangDyon,KaneDevice} where the
in-plane carrier transport would have a purely quantum topological
origin.

Topological insulators in two- and three-dimensions have been predicted
theoretically and observed experimentally as well to display quantum
spin Hall effect\cite{Kane1st,bernevig,BernevigSciHgTe} and strong
topological behavior due to the combined effects of relativistic
Dirac fermions and quantum-entanglement\cite{review,
Kane1st,FuKane,DavidNat1,KonigSci,DavidTunable,TIbasic,BernevigSciHgTe,Roy,MooreandBal}.
For systems with inversion symmetry, a band inversion occurs due
to a finite value of the spin-orbit coupling.
This results in a bulk-insulating gap which ensures the
existence of metallic surface states within the gap.\cite{ZhangPred}
In bismuth or antimony metals, numerous
surface states interfere with each other and an odd-number of Dirac
cones can be achieved by making composite alloys such as
Bi$_{1-x}$Sb$_x$.\cite{DavidNat1} In other novel classes of topological
insulators, the Dirac node appears either below the valence band
maximum as in Bi$_2$Te$_3$ or
Sb$_2$Te$_3$\cite{BiTeSbTe} or above the conduction
band minimum as in TlSbTe$_2$\cite{TlBiTe2}. Topologically nontrivial
materials such as
half-Heusler\cite{heuslerhasan} and Li$_2$AgSb\cite{HsinLi} series are
semi-metals, which require lattice distortion to be tuned into the
topological insulating state. For these reasons, it is
difficult to electrically gate these materials for the manipulation
and control of charge carriers for realizing a device. Therefore, a viable
route to search for the isolated Dirac node in topological insulating
materials is to find a chemically stable stoichiometric composition
in an appropriate class of materials.

With this motivation, we consider in this article five existing
stoichiometric tetradymite-like
ternary compounds in the structure $M_2X_2Y$, namely,
Bi$_2$Te$_2$Se\cite{Bi2Te2Se}, Bi$_2$Te$_2$S\cite{Bi2Te2S},
Bi$_2$Se$_2$S,\cite{Bi2Se2S} Sb$_2$Te$_2$Se,\cite{Sb2Te2Se} and
Sb$_2$Te$_2$S\cite{Sb2Te2S}. Tetradymite
structure is with a rhombohedral unit cell
of the space group $R\bar{3}m$. The commonly invoked hexagonal cell
shown in Fig.~1a
consists of three quintuple layers. The stacking order of two
consecutive units may be represented as
$-X-M-Y-M-X--X-M-Y-M-X-$. The layer of $Y$ atoms is in the center of the
unit. The natural surface
termination is between the two $X$-atom layers.
The known topological insulators Bi$_2$Se$_3$, Bi$_2$Te$_3$, and
Sb$_2$Te$_3$ have the same crystal structure with $X=Y$.

\begin{figure}
\includegraphics[width=8.5cm]{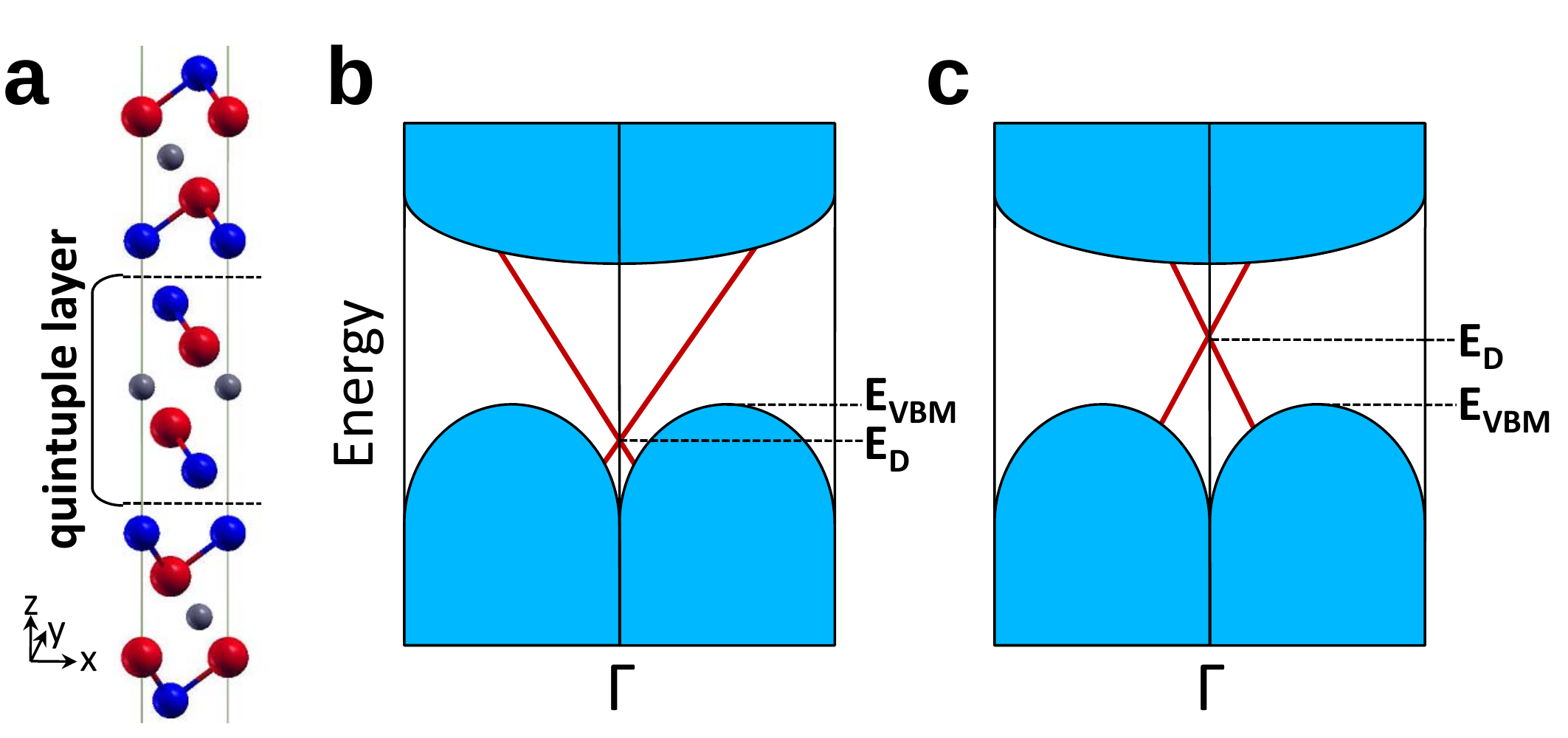}
\caption{\label{fig:sketch}
\textbf{Tetradymite crystal structure and topological transport regime}.
\textbf{a}, The crystal structure of tetradymite $M_2X_2Y$. $M$, $X$,
and $Y$ atoms are denoted by red, blue, and grey balls, respectively.
Diagrams in \textbf{b} and \textbf{c} illustrate surface band
structures near the $\Gamma$-point for stoichiometric Bi$_2$Te$_3$ and
Bi$_2$Se$_2$S.
}
\end{figure}

Fig.~1b shows a schematic illustration of the bulk and surface band
structures of Bi$_2$Se$_3$, Bi$_2$Te$_3$ and Sb$_2$Te$_3$ in which the
surface Dirac point lies below the
valence band maximum (VBM).
In our earlier experimental work, we lowered the Fermi level ($E_F$)
of Bi$_2$Se$_3$ into the bulk bandgap by substituting trace amount of
Ca$^{\rm 2+}$ for Bi$^{\rm 3+}$ with Ca acting as a hole doner to the
bulk states.
In order to lift the Dirac point above the VBM, we deposited NO$_2$ on
the surface. An isolated Dirac cone is then obtained as illustrated in
Fig.~1c.
In the present study, we predict new topological insulators
Bi$_2$Te$_2$Se, Bi$_2$Te$_2$S, Bi$_2$Se$_2$S, Sb$_2$Te$_2$Se
and Sb$_2$Te$_2$S, among which the latter three have a Dirac point
located in the bulk gap above the VBM as shown in Fig.~1c.
Therefore, these three compounds are well-suited as materials with an
isolated Dirac cone without requiring tuning via surface deposition.
With proper electron/hole doping control or gating, one can position
the $E_F$ at the Dirac point and the material can be in the transport
regime for potential device applications. We did not
attempt to tune the electronic structure of the studied compounds
using virtual crystal\cite{AB1,Lin1} or other
first-principles approaches\cite{AB2,AB3}.

\begin{figure}
\includegraphics[width=8.5cm]{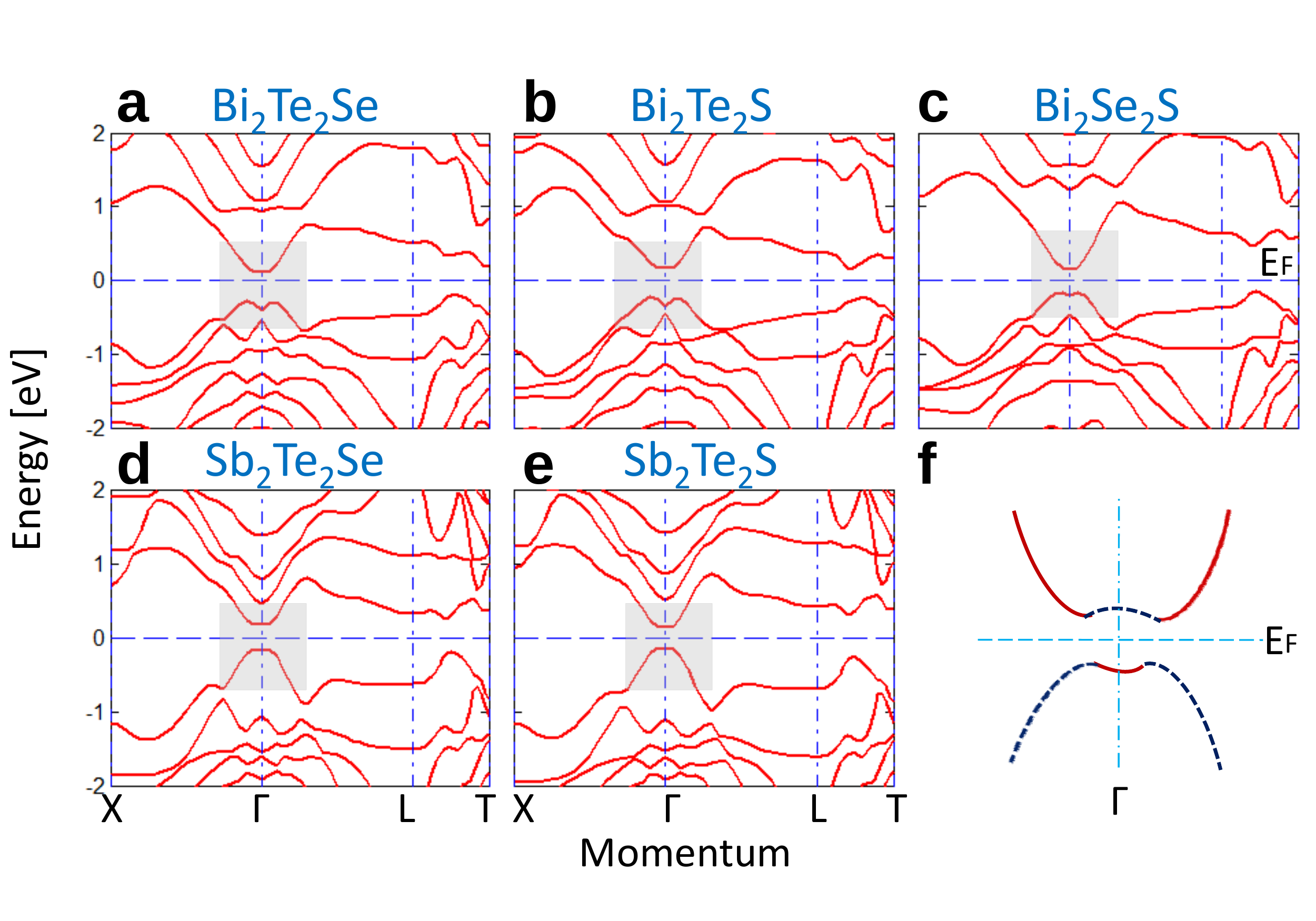}
\caption{\label{fig:band}
\textbf{Bulk electronic structures and band-inversion}.
Bulk electronic structures of  Bi$_2$Te$_2$Se (\textbf{a}),
Bi$_2$Te$_2$S (\textbf{b}), Bi$_2$Se$_2$S (\textbf{c}), Sb$_2$Te$_2$Se
(\textbf{d}) and Sb$_2$Te$_2$S (\textbf{e}). The grey shaded areas in
panels \textbf{a-e} highlight the inverted band structure as
illustrated in panel \textbf{f}.
}
\end{figure}

The first-principles band structures were computed on the
basis of experimental lattice data
\cite{Bi2Te2Se,Bi2Te2S,Bi2Se2S,Sb2Te2Se,Sb2Te2S} within the framework
of the density-functional theory (DFT) using the
generalized-gradient approximation (GGA). Our results, shown in
Fig.~2, predict that all these materials naturally
host a topological insulating phase with band gaps in the range of 224
meV to 297 meV.
A band inversion occurs at the
$\Gamma$-point as highlighted with grey shadings. As illustrated in
Fig.~2f, the conduction band (red solid line) and the valence band
(blue dashed line) cross each other near the $\Gamma$-point, so that
at this point the occupied state possess the character of the
conduction band and the unoccupied state that of the
valence band. Since the crystal has inversion symmetry, we have
applied band parity analysis \cite{FuKane} to
evaluate the value of the topological invariant $Z_2$. All five
compounds are thus found to be strong topological insulators with $Z_2=-1$.
This comes about due to band inversion at the $\Gamma$-point through
which the $Z_2$ topological invariant
picks up an extra factor of -1 compared to bands without SOC.
In short, the presence of an energy gap throughout the Brillouin
zone with a band inversion only at the $\Gamma$-point makes this
class of materials topological insulators.

\begin{figure}
\includegraphics[width=8.5cm]{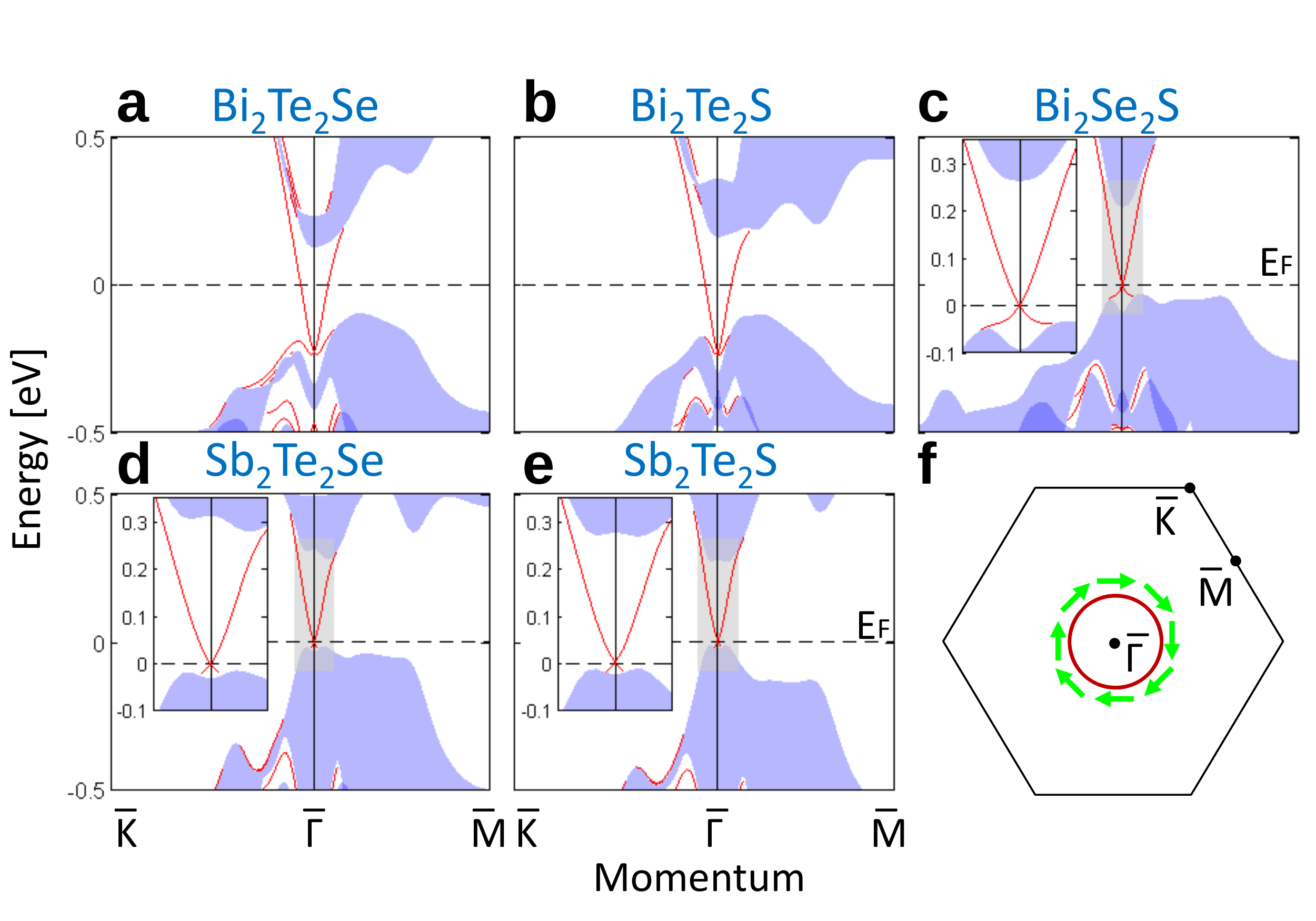}
\caption{\label{fig:surface}
\textbf{Surface states and spin-texture of the Dirac cone}.
Surface band structures of  Bi$_2$Te$_2$Se (\textbf{a}), Bi$_2$Te$_2$S
(\textbf{b}), Bi$_2$Se$_2$S (\textbf{c}), Sb$_2$Te$_2$Se (\textbf{d})
and Sb$_2$Te$_2$S (\textbf{e}). Red lines are the surface states
and the blue shaded areas are projected bulk states. Surface
momentum is defined in panel \textbf{f} showing the hexagonal surface
Brillouin zone. Green arrows in panel \textbf{f} denote the spin
direction of the
upper Dirac cone at constant energy (blue circle).
}
\end{figure}


The calculated DFT-GGA surface bands are shown in Fig.~3. A pair of
surface bands with opposite spin directions are degenerate at the
zone center and form the Dirac-cone-like dispersion. All five compounds
exhibit a single Dirac cone inside the bulk band gap.
For Bi$_2$Te$_2$Se and Bi$_2$Te$_2$S, the Dirac point lies below the
VBM. This is the same as the case of stoichiometric binary Bi$_2$Te$_3$
compound in which bulk scattering channels at the energy of the Dirac point
prevent interesting quantum phenomena in the topological transport
regime to be realized. The most interesting results are found in
Bi$_2$Se$_2$S, Sb$_2$Te$_2$Se, and Sb$_2$Te$_2$S where an isolated
Dirac cone is obtained.
Here, the Dirac point lies above the VBM and below the conduction bands.
When $E_F$ is tuned to the Dirac point, the density of states
approaches zero and there are no other scattering channels.
Therefore, Bi$_2$Se$_2$S, Sb$_2$Te$_2$Se, and Sb$_2$Te$_2$S would be
stoichiometric compounds which are wll-suited for investigating the
topological transport regime.

In conclusion, we predict that stoichiometric tetradymite-like
compounds Bi$_2$Te$_2$Se, Bi$_2$Te$_2$S,
Bi$_2$Se$_2$S, Sb$_2$Te$_2$Se, and Sb$_2$Te$_2$S are strong
topological insulators with large enough band gap for room temperature
applications.
The isolated Dirac cone on the surface of Bi$_2$Se$_2$S,
Sb$_2$Te$_2$Se, and Sb$_2$Te$_2$S will be particularly important for
studying topological transport properties
on a bulk material without the necessity of surface
deposition.\cite{DavidTunable} Experiments with magnetic scattering
techniques should be able to uniquely access the helical spin-texture of
the isolated surface states which are essential in device
applications. The stoichiometric crystals of the investigated
compounds are very easy to produce and manipulate for manufacturing
devices for microchips,\cite{microchip}
spintronics\cite{WolfSpintronics} and
information\cite{Majorana,LeekQC} technologies. Furthermore, the
full exposure of the topological transport regime for dissipationless
spin-current with the unique advantage of tunable surface states is
suitable for studying novel topological phenomenon such as intrinsic
quantum spin Hall effect,\cite{Kane1st,bernevig,BernevigSciHgTe}
universal topological magneto-electric effect\cite{essin} and
anomalous half-integer quantization of Hall
conductance\cite{TIbasic,TopoFieldTheory}, and the quantum
electrodynamical phenomena such as the image magnetic monopole induced
by an electric charge\cite{ZhangMonopole} or Majorana fermions
induced by the proximity effects from a superconductor\cite{Majorana}.

\textbf{Methods:}

First-principles band calculations were performed with the linear
augmented-plane-wave (LAPW) method using the WIEN2K package
\cite{wientwok} within the framework of the density functional theory (DFT).
The generalized gradient approximation (GGA)
was used to describe the exchange-correlation
potential.\cite{PBE} Spin orbital coupling (SOC) was included as a second
variational step using a basis of scalar relativistic eigenfunctions.
The surface was simulated by placing a slab of 6 quintuple layes for
Bi$_2$Te$_2$Se and Bi$_2$Te$_2$S.
For Bi$_2$Se$_2$S, Sb$_2$Te$_2$Se and Sb$_2$Te$_2$S we used 12
quintuple layers to ensure convergence with the number of layers.

\textbf{Acknowledgements:}
The work at Northeastern and Princeton is supported by the Division
of Materials Science and Engineering, Basic Energy Sciences, US
Department of Energy (DE-FG02-07ER46352, DE-FG-02-05ER46200 and
AC03-76SF00098), and benefited from the allocation of supercomputer
time at NERSC and Northeastern University's Advanced Scientific
Computation Center (ASCC). Support from the A. P. Sloan Foundation
(L.A.W., S.Y.X, and M.Z.H.) is acknowledged.


\end{document}